\documentclass[12pt]{article}

\usepackage{amsmath,amssymb}

\newcommand{\dd}{\partial}
\newcommand{\m}{\mu}
\newcommand{\n}{\nu}
\newcommand{\ls}{\left(}
\newcommand{\rs}{\right)}
\newcommand{\al}{\alpha}
\newcommand{\be}{\beta}
\newcommand{\ff}{\varphi}
\newcommand{\La}{\Lambda}
\newcommand{\te}{\theta}
\newcommand{\M}{{\mathcal M}}
\newcommand{\pp}{{\mathcal P}}
\newcommand{\pri}{{\,\in\,}}
\newcommand{\la}{\lambda}
\newcommand{\g}{\gamma}

\newcommand{\sh}{\sinh}
\newcommand{\ch}{\cosh}

\newcommand{\disn}[2]{$$\displaylines{\refstepcounter{equation}%
            \label{#1}\hskip 1em minus 1em #2\hfilneg}$$}
\newcommand{\nom}{\hfil\hskip 1em minus 1em (\theequation)}
\newcommand{\no}{\hfil \hskip 1em minus 1em\phantom{(\theequation)}%
            \hfilneg\cr\hfilneg\hskip 1em minus 1em\hfil}
\newcommand{\ns}{\hfill\cr\hfill}

\begin{document}

\title{
Embeddings for Schwarzschild metric:\\ classification and new results
}

\author{S.A.~Paston\thanks{E-mail: paston@pobox.spbu.ru},
A.A.~Sheykin\thanks{E-mail: anton.shejkin@gmail.com}\\
{\it Saint Petersburg State University, St.-Petersburg, Russia}
}
\date{\vskip 15mm}
\maketitle

\begin{abstract}
We suggest a method to search the embeddings of Riemannian spaces with a high enough symmetry in a flat ambient space.
It is based on a procedure of construction surfaces with a given symmetry.
The method is used to classify the embeddings of the Schwarzschild metric which have the symmetry of this solution,
and all such embeddings in a six-dimensional ambient space
(i.~e. a space with a minimal possible dimension) are constructed.
Four of the six possible embeddings are already known, while the two others are new.
One of the new embeddings is asymptotically flat, while the other embeddings in a six-dimensional
ambient space do not have this property.
The asymptotically flat embedding can be of use in the analysis of the many-body problem,
as well as for the development of gravity description as a theory of a surface in a flat ambient space.
\end{abstract}

\newpage

\section{Introduction}\label{p1}
It is known that every analytical four-dimensional space-time can be considered,
at least locally, as a surface in a flat space-time with $N\le 10$ dimensions,
if we consider the metric on the surface as being induced by a flat metric $\eta_{ab}$ of the ambient space.
This fact follows from the Janet-Cartan theorem \cite{gane,kart}
(it has been proven for the positive signature in the original works
and generalized for the arbitrary case in \cite{fridman61}).
It is convenient to describe the surface in terms of the embedding function $y^a(x)$
(where $\m=0,1,2,3$ and $a$ ranges over $N$ values),
in this case the induced metric is given by the formula
 \disn{vv1}{
g_{\m\n}(x)=\dd_\m y^a(x)\, \dd_\n y^b(x)\eta_{ab}
\nom}
(see for example \cite{eisenhart}).
A minimum value $N$ for which an embedding exists for a given space-time
(i.~e. there exists an embedding function $y^a(x)$ satisfying the equation (\ref{vv1}))
defines the "embedding class"{} $p=N-4$ of this space.
The higher is the symmetry of the space-time, the lower is its "embedding class", see~\cite{schmutzer}.

The problem of embeddings construction for various solutions of the Einstein equations is discussed since long ago, almost just after the appearance of General Relativity.
The aims of such construction can be different.
An explicit view of an embedding can be of use for better understanding of the space-time geometry,
this is especially clear on the example of the Fronsdal's embedding \cite{frons} for the Schwarzschild metric
(this embedding appears to be closely related to the use of Kruskal-Szekeres coordinates, see a remark at the end of the work \cite{frons}).
The existence of an embedding can be used in order to solve several problems being  not directly related to the embeddings, for example, in order to find the exact solutions of the Einstein equations, see~\cite{schmutzer}.
A possibility of embedding can also be used to formulate the gravity
under the form of a theory of a four-dimensional surface in a flat    ambient space  \cite{regge,deser}, similarly to the string theory formulation.

The problem of embeddings construction is discussed in many works.
An important result is the Kasner's theorem which states that a vacuum solution of the Einstein equations (corresponding to the  zero Einstein tensor) cannot be embedded in a five-dimensional ambient space,
i.~e. it has the "embedding class" higher than unity \cite{kasner2}.
Many concrete variants of embeddings
can be found in \cite{rosen65,collinson68},
and many of questions arising during the construction of embeddings are discussed in \cite{goenner}.
An extended bibliography related to the embedding theory and similar problems  can be found in \cite{tapiaob}.

One of the considered problems was the problem of embedding construction for the Schwar\-zschild solution.
Following the Kasner's theorem the ambient space dimension in this case $N\ge 6$.
The first embedding was also constructed by Kasner \cite{kasner3}, just  five years after the Schwarzschild solution appearance.
This embedding covers only the region outside the horizon, see details in Section \ref{p4var1}.
The problems of the global structure of a corresponding manifold are discussed in \cite{monte09}.
The most  global embedding covering all the regions corresponding to the Schwarzschild metric has been constructed by Fronsdal \cite{frons},
see Section \ref{p4var2} for its properties.
For both mentioned embeddings $N=6$, i.~e. the dimension of the ambient space is minimal.
For  $N=6$ two more embeddings are known, they have been suggested in  \cite{fudjitani} and \cite{davidson},
see details in Sections \ref{p4var5} and \ref{p4var3} respectively.
Note that in \cite{kasner3,frons,davidson} the embeddings were demonstrated without any method of their obtaining,
while in \cite{fudjitani} some method is used, but, for example, it does not give the  embedding \cite{davidson}.

In our work we suggest a method to search the embeddings based on the existing symmetry. The method works well if the embedded space-time has a high enough symmetry.
Note that in general case a surface being a some space-time embedding might have not all the symmetries of this space-time.
Such embeddings cannot be found by the method suggested below.
The idea of the method is that first we use the results of the groups representation theory in order to enumerate all the surfaces with the requested symmetry, and then to select among them those which have the required $N$ and metric.
The general method of the construction of surfaces with a given symmetry is described in Section \ref{p2}, where the used definition of the surface symmetry is also given.

We use the suggested method in order to construct the symmetrical embeddings of the  Schwarzschild metric.
In Section \ref{p3} we classify all the surfaces having the symmetry of the Schwarzschild solution,
and in Section \ref{p4} we give all of them with $N=6$ and the metric corresponding to this solution.
As a result we obtain all the symmetric embeddings of the Schwarzschild metric in a space of minimal dimension.
It appears that there exist six types of such embeddings, four of them coincide with the already known embeddings mentioned above,
while the two others (see Sections \ref{p4var4} and \ref{p4var6} for their properties) have never been described, as far as we know.

Only one of these six embeddings is asymptotically flat
(i.~e. the surface corresponding to it tends to a four-dimensional plane while moving off the center),
this is the new embedding from the section \ref{p4var4}.
The asymptotically flat embedding is of use in the analysis of the many-body problem.
Moreover, the existence of the asymptotically flat embedding is of great importance for the development of the approach suggested in \cite{regge},
where the gravitation is described not by the metric, but by the embedding function defining a four-dimensional surface in a flat space.
Indeed, it should be reasonable to suggest that while moving far off the gravitating body the surface must tend to some "average"{} surface,
being homogeneous  and isotropous over the scales small with respect to the scale of the Universe (where in the framework of the
Friedmann-Robertson-Walker (FRW) model
the four-dimensional homogeneity  and isotropy are violated).
And while over  these scales the average curvature of the space-time is not noticeable,
one can consider that the "average"{} surface must be a four-dimensional plain,
and hence the embedding of the gravitating body metric must be asymptotically flat.
The examples of the approach \cite{regge} development are the canonical formulation of the theory and the discussion of its quantization \cite{davkar},
the canonical formulation with  an additional imposition of constraints suggested in \cite{regge}
which ensure the absence of extra solutions \cite{tmf07,ijtp10},
the transition to the field theory in a flat ambient space \cite{tmf2011},
as well as other  works \cite{pavsic85let,tapia,maia89,bandos}, see also the overview~\cite{tapiaob}.

Note that there exists an asymptotically flat embedding of the Schwarzschild metric, suggested in \cite{blashke2010},
but it is an embedding in a  seven-dimensional space with two time-like directions. In contrast, in Section \ref{p4var4} an asymptotically flat embedding in a six-dimensional space with one time-like direction is constructed.

\section{Construction of surfaces with a given symmetry}\label{p2}
Let us construct a $d$-dimensional surface in a flat space symmetrical
with respect to a given group $G$. In a general case a flat space contains $n_+$ time-like and $n_-$ space-like directions, $n_++n_-=N>d$,
i.~e. it is a $R^{n_+,n_-}$ space.
Suppose that the surface $\M$ is symmetrical with respect to a group $G$, if $\M$ passes into itself under the action of some  sub-group of the group of motions $\pp$ of the flat space $R^{n_+,n_-}$, when this sub-group is isomorphous to $G$.
For $R^{n_+,n_-}$ the group of motion is a corresponding generalization of the Poincar\'e group,
i.~e. $\pp$ is a semidirect product of the group $SO(n_+,n_-)$ by the group  $T^{n_++n_-}$ of translations of the space $R^{n_+,n_-}$.
The used definition of symmetry corresponds to the fact that the regions passing into one another under the symmetry transformations must have the same internal and external geometry,
hence they can be superimposed one to another under the action of shift and (or) rotation in the ambient space.

Hence in order to construct the required surface we  need to find some homomorphism $V$ from the group $G$ into the group $\pp$.
As we know, the elements $\pp$ can be represented under the form of block matrice of the size $n_++n_-+1$ having the form
 \disn{0.1}{
\ls
\begin{array}{cc}
\La & a\\
0   & 1\\
\end{array}
\rs,
\nom}
where $\La\pri SO(n_+,n_-)$ and $a\pri R^{n_+,n_-}$ parametrize the translation
(while the points $R^{n_+,n_-}$ correspond to $n_++n_-+1$-dimensional vectors with the last component always equal to unity).
Hence $V$ can always be considered as a representation of the $G$ group whose matrices have the form (\ref{0.1}),
and the representation must be single-valued and exact, but it will be reducible
(because the matrix (\ref{0.1}) corresponds to a reducible representation, although not completely  reducible).
Then the  set of points $y\pri R^{n_+,n_-}$ given by the formula
 \disn{1}{
y=V(g)y_0
\nom}
at an arbitrary $g\pri G$ and a fixed initial vector $y_0\pri R^{n_+,n_-}$
will result,  by construction, in a surface with the required symmetry.
If the vector $y_0$ will be constructed as dependent on some continuous parameters,
then the  set of points given by (\ref{1}) for all values of these parameters will also form a surface with the required symmetry,
and in this case its cross-sections corresponding to the fixed values of these parameters will have the same symmetry.

Since the classification of group representations is well developed,
a constructive way of  constructing the surfaces with the required symmetry can be suggested:
to enumerate all real representations $V$ of the group $G$ starting from their minimal dimensions,
and to select the representations with the representation matrix having the form (\ref{0.1}).
One must however take into account that for different representations and different initial vectors $y_0$ the obtained surfaces can have different dimensions,
hence not all variants will be suitable at a given surface dimension $d$.

Note that depending on the problem to be solved the requirement for the symmetry of the desired surface can be formulated either globally or locally. In the last case the sub-group of the group of motions of the ambient space which leaves invariant the surface $\M$ must be only locally isomorphous to the group $G$. Then the representation $V$ needs not to be single-valued, and it can have a non trivial discrete kernel.
Such situation can take place if we search for an embedding for the Riemannian space
whose symmetry is observed only locally but not globally.
For example, the closed  FRW Universe has the $SO(4)$ symmetry, but we cannot affirm it globally, since we cannot pass it around over the great circle.
On the other hand, in some tasks it can be necessary not only to suppose a presence of a global symmetry, but to impose additional limitations on the topology of the desired surface.
For example, when we look for an embedding for the Schwarzschild metric discussed below,
since the gravitating body can be passed around over a great circle,
we have to consider the symmetry $SO(3)$ as global, and to require in addition the topology of the surface corresponding to the constant values $r$ and $t$ to be the sphere topology.

\section{Classification of surfaces with the symmetry\\ of Schwarzschild metric}\label{p3}
Let us apply the method of construction of surfaces with a given group of symmetry described in the previous section in order to find various embeddings of the Schwarzschild metric.
The Schwarzschild metric represents a spherically symmetric solution of the Einstein equations in the absence of matter.
The interval corresponding to this metric reads
 \disn{2}{
ds^2=\ls 1-\frac{R}{r}\rs dt^2-\frac{dr^2}{1-\frac{R}{r}}
-r^2\ls d\te^2+\sin^2\te\, d\ff^2\rs,
\nom}
where $R$ is the Schwarzschild radius. The Schwarzschild metric is also invariant with respect to the shifts of the coordinate $t$,
hence the full symmetry of the  space-time with the Schwarzschild metric is described by the group $G=SO(3)\times T^1$.

Let us look for the general form of a four-dimensional surface $\M$
symmetrical with respect to the given group $G$.
We will suppose that not only the whole four-dimensional surface $\M$ must have this symmetry,
but also all its sub-manifolds corresponding to the fixed values of the parameter $r$.
Stress that, as already mentioned in the Introduction,
a surface being an embedding of some space-time might have not all the symmetries of this space-time,
but we will not consider here these variants.
For the convenience the main results obtained in this Section are gathered in the end of the Section.

The symmetry of $\M$ with respect to $G$ means that $\M$ passes into itself under the action of $V(g)\pri\pp$,
where $g\pri G$, and $V$ is a representation of the group $G$.
Consider a possible form of this representation.
The representations $V$ of the direct product $SO(3)\times T^1$ are reduced to the direct sum of the representations $\tilde V$,
being the tensor products of the representations of groups $SO(3)$ and $T^1$:
 \disn{z2}{
\tilde V(g)= \tilde V_1(O)\otimes \tilde V_2(t),\qquad
O\pri SO(3),\quad t\pri T^1,
\nom}
and we can  suppose that the representations $\tilde V_1$ and $\tilde V_2$ are not  completely reducible.

Consider first the representations of the group $SO(3)$.
It is known that all its finite-dimensional irreducible representations can be obtained from
tensor representations of an universal covering group $SU(2)$ classified by their spin.
Since we suppose that the symmetry with respect to $G$ is inherent to sub-manifolds $\M$ corresponding to the fixed values of $r$,
the surface formed by the points satisfying (\ref{1}) at $g=\ls O\times {\bf 1}\rs$
corresponds to fixed values of $r$ and $t$. This is why it has to be two-dimensional, since two more dimensions are associated to the  variation of these parameters.
Since the elements of the group $SU(2)$ depend on three real parameters,
that imposes a limitation on the choice of the representation as well as of the initial vector $y_0$.
Namely, in $SO(3)$ there must exist a single-parametric subgroup $SO(2)$ for whose elements $V(g)y_0=y_0$, i.~e. a one-dimensional stability subgroup of the vector $y_0$.
One can show that for all representations with a half-integer spin
(acting in the real spaces which are obtained as the realification of corresponding complex spaces)
this condition cannot be satisfied.
This is why we have to limit ourselves by the irreducible representations with an integer spin
which can be considered as tensor representations of the group $SO(3)$
acting in the spaces of real irreducible tensors $A_{i_1\ldots i_m}$ ($i,k=1,2,3$).
In this case the mentioned condition means that the tensor $A^0_{i_1\ldots i_m}$ corresponding to the initial vector $y_0$ must be invariant with respect to the rotations around some axis,
hence it must be expressed via a vector $x_i$ directed along this axis:
 \disn{z13}{
A^0_{i_1\ldots i_m}=\ls x_{i_1}\ldots x_{i_m}-\ldots\rs,
\nom}
where the ellipsis means the terms ensuring the irreducibility of the given tensor,
and $x_i$ is the vector being transformed accordingly to the vector representation of the group $SO(3)$.

Hence the representation space for $\tilde V_1$ from the eq. (\ref{z2})
is a space of irreducible symmetric tensors $A_{i_1\ldots i_m}$
and is characterized by an integer number $m\ge 0$,
while the component $C^0$ of the initial vector $y_0$ related to the space of the representation $\tilde V$ is
 \disn{3}{
C^0=\ls A^0\otimes B^0\rs,
\nom}
where $A^0$ is given by the formula (\ref{z13}),
and $B^0$ belongs to the space of the representation $\tilde V_2(t)$ and will be concretized below. In order to stress out that the representation $\tilde V_1$ is characterized by an integer number $m\ge 0$, we will write below $\tilde V_1^m$ instead $\tilde V_1$.

As already mentioned in the end of Section~\ref{p2}, in the considered problem
we have to consider the symmetry $SO(3)$ as global, and to require in addition the topology of the  two-dimensional surface to be corresponding to fixed values of the parameters $r$ and $t$ to be the sphere topology.
The first one is also a reason why we must reject the double-valued representations corresponding to a half-integer spin
(they have been excluded above as giving a too big dimension of the surface).
The second one leads to the existence of at least one odd rank $m$ in the direct sum defining the full space where the representation $V$ acts.
Indeed, if all the values $m$ are even, then to the vectors $x_i$ and $-x_i$ will correspond one and the same point of the constructed surface,
and its topology will be the topology of a sphere with identified opposite points, hence different from the sphere topology.

Consider now the representations $\tilde V_2$ of the  abelian group $T^1$ in the formula  (\ref{z2}).
Assume that the group elements are real numbers, and the group operation is the addition.
Then in the most general case the representation matrix $\tilde V_2(t)$ can be written under the form
 \disn{z15}{
\tilde V_2(t)=e^{\g tW},
\nom}
where $W$ is an arbitrary fixed complex matrix and $\g$ a positive dimensional factor,
introduced in order to make $W$ dimensionless.
We can consider that the representation $\tilde V_2$ is not completely  reducible. Hence, since any matrix can be reduced to a Jordan form by a basis choice,
$W$ can be written under the form of a matrix of size $s+1$ composed of one Jordan block:
 \disn{z16}{
W=\ls
\begin{array}{cccc}
\la    & 1      & 0      & \dots \\
0      & \la    & 1      & \dots \\
\vdots  & \vdots  & \ddots  & \ddots \\
0      & 0      & 0      & \la   \\
\end{array}
\rs,
\nom}
where $\la$ is a complex number. Writing it down under the form $\la=(\be+i\al)/\g$ and substituting it (\ref{z16}) into (\ref{z15}), we obtain
 \disn{z17}{
\tilde V_2(t)=S^s(\g t)e^{i\al t}e^{\be t},
\nom}
where $S^s(\g t)$ is a matrix of size $s+1$ having the form
 \disn{z17.1}{
S^1(z)=
\ls
\begin{array}{cc}
1      & z\\
0      & 1\\
\end{array}
\rs,
\quad
S^2(z)=
\ls
\begin{array}{ccc}
1      & z      & \frac{z^2}{2!}\\
0      & 1      & z\\
0      & 0      & 1\\
\end{array}
\rs,
\quad
S^3(z)=
\ls
\begin{array}{cccc}
1      & z      & \frac{z^2}{2!} & \frac{z^3}{3!}\\
0      & 1      & z              & \frac{z^2}{2!}\\
0      & 0      & 1              & z\\
0      & 0      & 0              & 1\\
\end{array}
\rs,
\quad
\ldots
\nom}

Now we introduce for representation $\tilde V_2$ the number $p,q=0,1$ in order to have
$p=0$  if$\al=0$, otherwise $p=1$;
and $q=0$  if $\be=0$, otherwise $q=1$.
The formula (\ref{z17}) can be interpreted as an expression for the representation $\tilde V_2$
under the form of a tensor product of three representations,
characterized by the numbers $s\ge0$, $p,q=0,1$ and real nonzero factors $\g$, $\al$, $\be$.
Taking into account (\ref{z2}), it allows to write the representation $\tilde V$ under the form
 \disn{z20}{
\tilde V(g)=\tilde V_1^m(O)\otimes S^s(\g t)\otimes P^p(\al t)\otimes Q^q(\be t),
\nom}
where $P^1(z)=e^{iz}$, $Q^1(z)=e^{qz}$, $S^0(z)=P^0(z)=Q^0(z)={\bf 1}$,
and to characterize it by a set of numbers $\{m,s,p,q\}$ and of dimensional factors $\g$, $\al$, $\be$.

According to what is written in Sec.~\ref{p2} (see after the formula (\ref{1})),
we have now to  make the realification of the space of the representation $P^1$
(the remaining tensor factors in (\ref{z20}) already act in real spaces).
As a result we have
 \disn{z19}{
P^1(\al t)=\ls
\begin{array}{cc}
\cos\al t   & -\sin\al t     \\
\sin\al t   & \cos\al t     \\
\end{array}
\rs,
\nom}
i.~e. $P^1(\al t)$ is an orthogonal matrix.
The arbitrariness in the basis choice allows to consider $\al>0$.

Then we have to choose a set of representations $\tilde V$ under the form (\ref{z20}), in order to be able to write the matrix of representation for their direct sum $V$ (see text before the formula (\ref{z2})) under the form (\ref{0.1}).
It means in particular that in the subspace with the dimension being lower for a unity than the dimension of the whole representation space
these matrices must be orthogonal or pseudo-orthogonal.

In order to satisfy this condition we must reduplicate the dimension of the space of the representation $Q^1$,
having taken the direct sum of one-dimensional representations with different signs $\be$ (we can consider in the future that $\be>0$), i.~e. to take
 \disn{z21}{
Q^1(\be t)=\ls
\begin{array}{cc}
e^{\be t}   & 0     \\
0   & e^{-\be t}     \\
\end{array}
\rs.
\nom}
It is easy to demonstrate a quadratic form which is invariant with respect to such representation:
 \disn{z21a1}{
\eta=\ls
\begin{array}{cc}
0   & 1     \\
1   & 0     \\
\end{array}
\rs
\qquad\Rightarrow\qquad
Q^1(\be t)\,\eta\, (Q^1(\be t))^t=\eta.
\nom}
It is well known that there exist a basis where the values $\eta$ and $Q^1(\be t)$ take the form
 \disn{z21a2}{
\eta=\ls
\begin{array}{cc}
1   & 0     \\
0   & -1     \\
\end{array}
\rs,\qquad
Q^1(\be t)=\ls
\begin{array}{cc}
\ch\be t   & \sh\be t     \\
\sh\be t   & \ch\be t     \\
\end{array}
\rs,
\nom}
i.~e. $Q^1(\be t)$ appears to be a pseudoorthogonal matrix.
Note that the basis where the quadratic form has the structure (\ref{z21a1}) corresponds to the use of light-like coordinates in a pseudoeuclidian space.

Consider now the matrices $S^s(\g t)$. One can show that the transformations given by these matrices leave invariant the quadratic form $\eta^s$ having the structure
 \disn{z21a3}{
\eta^1=
\ls
\begin{array}{cc}
0      & 1\\
-1      & 0\\
\end{array}
\rs,
\quad
\eta^2=
\ls
\begin{array}{ccc}
0      & 0      & 1\\
0      & -1     & 0\\
1      & 0      & 0\\
\end{array}
\rs,
\quad
\eta^3=
\ls
\begin{array}{cccc}
0  & 0      & 0      & 1\\
0  & 0      & -1     & 0\\
0  & 1      & 0      & 0\\
-1 & 0      & 0      & 0\\
\end{array}
\rs,
\quad
\ldots
\nom}
containing alternate numbers $+1$ and $-1$ on the anti-diagonal.
For even $s$ this quadratic form is symmetric.
Similarly to the case of the  two-dimensional quadratic form (\ref{z21a1}), it can be transformed  in this case to a standard diagonal form of the metric of space $R^{s/2,1+s/2}$ passing from light-like coordinates to Lorenz ones.
This is why for even values $s$ the matrices $S^s(\g t)$ are pseudo-orthogonal in some basis.

On the other hand, for odd $s$ one can show that there exist no invariant non-degenerated symmetric quadratic form,
this is why the matrix $S^s(\g t)$ is not orthogonal or pseudoorthogonal.
But we can note that in this case the matrix $S^s(\g t)$ has the form (\ref{0.1}),
and its part corresponding to the value $\La$ from (\ref{0.1}) coincides with $S^{s-1}(\g t)$,
hence in some basis it is a pseudoorthogonal matrix.

This is why odd values, as well as even values of $s$ are possible,
but the variant with odd $s$ is allowed only for one of the terms $\tilde V$ constituting $V$, and so only if $m=p=q=0$
(otherwise the matrix of the representation $V$ cannot be written under the form (\ref{0.1})).

As a result for odd values $s$ ($s=2j+1$) the space of the representation $\tilde V$ is characterized by a set of numbers
$\{0,2j+1,0,0\}$
(this variant can occur no more than in one term in $V$) and
it is a space of $(2j+2)$-dimensional vectors $C_F$,
for which the action $\tilde V$ having the form $C'=\tilde V(g)C$ is written as
 \disn{z24}{
C'_F=S^{2j+1}_{FG}(\g t)\,C_{G},
\nom}
where $F,G=1,\ldots,2j+2$.
If $j=0$ (i.~e. $s=1$), then it follows from (\ref{z24}) that in such case the translational invariance with respect to time
is  realized by shifts along some direction in the ambient space.

For even values $s$ ($s=2j$) the space of the representation $\tilde V$ is characterized by a set of numbers
$\{m,2j,p,q\}$, and at $p=q=1$ it appears to be a space of tensors
$C_{i_1\ldots i_m FAC}$, where
 \disn{z24a1}{
C'_{i_1\ldots i_m FAC}=O_{i_1k_1}\ldots O_{i_mk_m}S^{2j}_{FG}(\g t)P^1_{AB}(\al t)Q^1_{CD}(\be t)\,C_{k_1\ldots k_m GBD},
\nom}
and $A,B,C,D=1,2$.
If a number  $p$ or $q$ vanishes, then $C$ looses the corresponding index $A$ or $C$.
In the case under consideration the translational invariance with respect to time
is realized  under the form of combinations of rotations in the ambient space
(they can be usual rotations, Lorenz boosts or rotations in the light-like space).
There can be any number of terms in  $V$ corresponding to this case,
but at least in one of them the number $m$ must be odd (see text after the formula (\ref{3})),
and at least in one of them the numbers $s$, $p$ or $q$ must differ from zero,
if the case $\{0,2j+1,0,0\}$ has not been realized in any of the terms in $V$.
The last condition is a consequence of the requirement of the absence of the kernel for the representation $V$,
since if it is violated, then it will be $V({\bf 1}\times t)={\bf 1}$.

Let us found now a possible form of the initial vector $y_0$ in Eq. (\ref{1}).
It has been shown earlier that its projection on the representation space $\tilde V$ has the form (\ref{3}).
For the case $\{0,2j+1,0,0\}$ this projection is a $(2j+2)$-dimensional vector under the form
 \disn{z25}{
C^0_F=\ls
\begin{array}{c}
h \\
v \\
\end{array}
\rs\equiv D^{2j+2}_F,
\nom}
where the last component is equal to a dimensional constant $v$ and can be transformed into a unity by multiplication of the last basis element by $v$.
As a result the condition given in brackets after the formula (\ref{0.1}) will be satisfied.
In this case the last column of the matrix $S^{2j+1}(\g t)$ will be multiplied by $v$ and will become dimensional.

For the case $\{m,2j,p,q\}$, following (\ref{3}) one can write
 \disn{z26}{
C^0_{i_1\ldots i_m FAC}=A^0_{i_1\ldots i_m}B^0_{FAC},
\nom}
where $A^0_{i_1\ldots i_m}$ is given by the formula (\ref{z13}), and $B^0_{FAC}$ is arbitrary.
It is important to stress out that the direction of the vector $x_i$ by which the value $A^0_{i_1\ldots i_m}$ is defined
must be the same for all terms $V$,
since it is defined by a stability sub-group of the vector $y_0$,
see the text after the formula (\ref{z2}).

Since we suppose that not only the whole four-dimensional surface $\M$,
but also all its three-dimensional sub-manifolds corresponding to fixed values of the parameter $r$ must be symmetric with respect to $G$,
then the initial vector $y_0$ must depend on $r$,
see the text after the formula (\ref{1}).
The arbitrariness in the choice of this dependence is limited by the requirement imposed to the value $C^0$ by the formulas (\ref{z25}) and (\ref{z26}).
At that the constant $v$ cannot depend on $r$,
because there must be a possibility to  transform it into the unity by a choice of the basis,
and the direction of the vector $x_i$ is defined by the stability sub-group of the vector $y_0$, hence it also cannot depend on $r$.
The vector $x_i$ can be considered, without restriction of generality, as normalized by the unity,
since in the expression (\ref{z26}) the change of its length can be compensated by a rescaling of the value $B^0_{FAC}$.
Then $x_i$ becomes a constant vector since its direction is fixed (see above).
As a result we can conclude that the whole dependence of the initial vector $y_0$ on $r$ reduces to the fact that the values $h$ from (\ref{z25}) and $B^0_{FAC}$ from (\ref{z26}) are the arbitrary functions of $r$.
In each term of the direct sum in $V$ these functions can be selected independently.

For convenience of reading we state again the results obtained in the present Section.
All the four-dimensional surfaces $\M$ in the space $R^{n_+,n_-}$ with the symmetry of the Schwarzschild solution can be obtained as an  set of points
 \disn{4}{
y=V(g)y_0, \qquad
g\pri SO(3)\times T^1,
\nom}
where $V$ is a representation being a direct sum of the representations $\tilde V$ of the form (\ref{z20}), and $y_0$ being the initial vector.
One of the terms of this sum can be a representation characterized by a set of numbers $\{0,s,0,0\}$, $s=2j+1$,
this representation being defined by the formula (\ref{z24}).
The projection of the initial vector $y_0$ on the space of such representation is given by the formula (\ref{z25})
(see also the remark after this formula),
the value $h$ in this formula depends on $r$.
The remaining terms of this sum are the representations characterized by sets of numbers $\{m,s,p,q\}$, $s=2j$,
defined by the formula (\ref{z24a1}).
The projection of the initial vector $y_0$ on the space of such representation is given by the formula (\ref{z26})
where the value $B^0_{FAC}$ depends on $r$,
while $A^0_{i_1\ldots i_m}$ is expressed via a constant unit vector $x_i$ by the formula (\ref{z13}).
At least one of these representations must contain an odd number $m$, and at least one of them must contain a non-zero $s$, $p$ or $q$.

It is easy to find the contribution $\tilde N$ of a concrete  representation $\tilde V$ in the value of the full dimension $N$ of the ambient space which is the sum of all such contributions.
For $\{0,2j+1,0,0\}$ it is clear that
\disn{5.1}{
\tilde N=s=2j+1,
\nom}
taking into account that the last component of the vector (\ref{z25}) must not be considered,
since its value is always fixed.
Concerning $\{m,2j,p,q\}$, we can conclude from the structure of the indexes of the tensor $C_{i_1\ldots i_m FAC}$ that
\disn{5.2}{
\tilde N=(2m+1)\,(s+1)\,2^{p+q}=(2m+1)\,(2j+1)\,2^{p+q},
\nom}
where we used the fact that the dimension of the space of three-dimensional irreducible tensors of the rank $m$ is equal to $2m+1$.

\section{Embedding of the Schwarzschild metric\\ in a six-dimensional space}\label{p4}
\subsection{General analysis}\label{p4ob}
As we mentioned in the Introduction, the minimal dimension $N$ of the flat space
in which the Schwarzschild metric can be embedded is equal to six.
Using the classification of surfaces with the symmetry of the Schwarzschild metric obtained in the previous Section  we will enumerate all the possible symmetric embeddings of this metric
in six-dimensional ambient spaces. To do this we will enumerate all the symmetric surfaces for $N=6$, and for each of them we will verify a possibility to represent its metric under the form (\ref{2}) using the induced metric formula (\ref{vv1}).

First of all, let us consider the limitations imposed by the condition $N=6$ on the values of the parameter $m$, which characterizes possible representations of $\tilde V$.
Using the formula (\ref{5.2}) and the fact that at least one odd number $m$ and a non-zero $s$, $p$ or $q$ must be present,
we can conclude that in the direct sum defining the representation $V$ only the terms $\tilde V$ with $m=1$ (once) and with $m=0$ can be found.
 At that for the term with $m=1$ there must be $s=0$,
while $p$ and $q$ are either both equal to zero, or one of them is equal to zero, and another one is equal to unity.

In the last case two possibilities arise for which the representation $V$ is written as
 \disn{6}{
V_1=\{1,0,0,1\},\qquad
V_2=\{1,0,1,0\},
\nom}
where the set of numbers $\{\ldots\}$ here and below denotes the representation $\tilde V$ corresponding to this set.
According to the formulas (\ref{z26}),(\ref{z13}) for the variant $V_1$ the initial vector $y_0$ is $x_i B^0_C(r)$.
Since $x_i$ is a fixed unit vector, the result of the action of the orthogonal matrix $O$ on it
can be written under the form
 \disn{7}{
O_{ik}x_k=\{\cos\te,\,\,\sin\te\,\cos\ff,\,\,\sin\te\,\sin\ff\}.
\nom}
Using such parametrization the embedding function corresponding to $V_1$ and constructed following (\ref{z24a1}),(\ref{z21a2}),
can be written for some of values $r$ as
 \disn{8}{
\begin{array}{lcl}
y^0=f(r)\sh(\be t+w(r))\,\cos\te,                   &\qquad &  y^3=f(r)\ch(\be t+w(r))\,\cos\te,  \\
y^1=f(r)\sh(\be t+w(r))\,\sin\te\,\cos\ff,          &\qquad &  y^4=f(r)\ch(\be t+w(r))\,\sin\te\,\cos\ff,\\
y^2=f(r)\sh(\be t+w(r))\,\sin\te\,\sin\ff,          &\qquad &  y^5=f(r)\ch(\be t+w(r))\,\sin\te\,\sin\ff,\\
\end{array}
\nom}
while for other values $r$ it can be written in a similar form, but with the interchange of hyperbolic sine and cosine
(depending on the sign of the square of the vector $B^0_C(r)$).
In this case the signature of the ambient space can be unambiguously chosen as $(+++---)$.
For the variant $V_2$ the embedding function is similar (\ref{8}), but there are normal functions instead of hyperbolic ones,
and the signature corresponds to the Euclidean space.

We cannot obtain the Schwarzschild metric embedding under the form of a surface corresponding to $V_1$ or $V_2$,
as the number of arbitrary functions is too small
(for example, for $V_1$ using the formula (\ref{vv1}) we found from the requirement $g_{22}=-r^2$  that $f(r)=r$,
after what the form $g_{00}$ corresponding to (\ref{2}) cannot be obtained any more).
Hence we can conclude that for the Schwarzschild metric  we have
 \disn{9}{
V=\{1,0,0,0\} \oplus W,
\nom}
where $W$ is the direct sum of the representations $\tilde V$ with $m=0$, and their total dimension must be equal to three.
Therefore, using (\ref{7}) and satisfying immediately the condition $g_{22}=-r^2$, we can write the embedding function as
 \disn{11}{
\begin{array}{lcl}
y^0=y^0(t,r),          &\qquad &  y^3=r\,\cos\te,  \\
y^1=y^1(t,r),          &\qquad &  y^4=r\,\sin\te\,\cos\ff,\\
y^2=y^2(t,r),          &\qquad &  y^5=r\,\sin\te\,\sin\ff,\\
\end{array}
\nom}
and the signature of the ambient space must ensure that $y^3,y^4,y^5$ are the space-like directions.

Consider first the cases when $s=0$ in all the terms in $W$.
Using the formula (\ref{5.2}) we write down all the variants giving the dimension three:
 \disn{12}{
W_1=\{0,0,1,0\}\oplus \{0,0,0,0\},\qquad
W_2=\{0,0,0,1\}\oplus \{0,0,0,0\}.
\nom}

\subsection{Variant 1: Kasner's embedding}\label{p4var1}
Consider the variant $W_1$. According to (\ref{z26}), in this case the initial vector $y_0$ is $B^0_A(r)\oplus B^0(r)$,
resulting (following (\ref{z24a1}),(\ref{z19})) in the embedding function
 \disn{13}{
y^0=f(r)\sin(\al t+w(r)),\quad
y^1=f(r)\cos(\al t+w(r)),\quad
y^2=g(r)
\nom}
(the three other components are always given by (\ref{11})).
Substituting it in (\ref{vv1}) with the right side corresponding to (\ref{2}), for $r>R$ we found that
the signature must be $(++----)$, and the embedding function becomes
 \disn{14}{
y^0=\frac{1}{\al}\sqrt{1-\frac{R}{r}}\,\sin(\al t),\quad
y^1=\frac{1}{\al}\sqrt{1-\frac{R}{r}}\,\cos(\al t),\quad
y^2=\int\! dr \sqrt{\frac{R(R+4\al^2 r^3)}{4\al^2 r^3(r-R)}}.
\nom}

This embedding was found in 1921 in \cite{kasner3} (it is supposed there $\al=1$)
and historically it is the first Schwarzschild metric embedding.
It covers only the region $r>R$ and has a conic singularity at $r=R$.
This embedding is not asymptotically flat, i.~e. it does not tend to a four-dimensional plane at
$r\to\infty$, and it cannot be described in terms of elementary functions.
Note that if we pass to the signature $(--+---)$, the form (\ref{13}) can also be used
to  construct an embedding of the Schwarzschild metric at $r<R$.

\subsection{Variant 2: Fronsdal's embedding}\label{p4var2}
Consider now the variant $W_2$.
In this case according to (\ref{z26}) the initial vector is $B^0_C(r)\oplus B^0(r)$,
resulting in the embedding function which, according to (\ref{z24a1}),(\ref{z21a2}), for some values of $r$ can be written under the form
 \disn{15}{
y^0=f(r)\sh(\be t+w(r)),\quad
y^1=f(r)\ch(\be t+w(r)),\quad
y^2=g(r)
\nom}
(the three other components follow the eq. (\ref{11})).
For other values of $r$ it can be written under a similar form, but with the interchange of hyperbolic sine and cosine
(depending on the sign of the square of the vector $B^0_C(r)$).
Substituting this embedding function into (\ref{vv1}),(\ref{2}), we found that
for the signature $(+-----)$ the embedding function becomes
 \disn{16}{
r>R:\quad
y^0=\frac{1}{\be}\sqrt{1-\frac{R}{r}}\,\sh(\be t),\quad
y^1=\pm\frac{1}{\be}\sqrt{1-\frac{R}{r}}\,\ch(\be t),\quad
y^2=g(r),\no
r<R:\quad
y^0=\pm\frac{1}{\be}\sqrt{\frac{R}{r}-1}\,\ch(\be t),\quad
y^1=\frac{1}{\be}\sqrt{\frac{R}{r}-1}\,\sh(\be t),\quad
y^2=g(r),
\nom}
where the function $g(r)$ at $\be=1/(2R)$ is well defined for all $r$ and smooth in the point $r=R$:
 \disn{17}{
g(r)=\int\! dr \sqrt{\frac{R(R-4\be^2 r^3)}{4\be^2 r^3(R-r)}}=
\int\! dr \sqrt{\frac{R}{r}+\ls\frac{R}{r}\rs^2+\ls\frac{R}{r}\rs^3}.
\nom}

This embedding is possibly the most well known, it was suggested in \cite{frons}.
The corresponding manifold is smooth everywhere including $r=R$
 (to make sure of it we must describe it using two equations for $y^a$, see~\cite{frons}),
and for $r\to0$ we have $y^0\to\pm\infty$, i.~e. the central singularity appears to be at the infinity.
This embedding differs from the other ones by the fact that it covers all the regions of the Riemannian space corresponding to the Schwarzschild solution: two instances of the region $r>R$ and two instances of the region $r<R$,
related to a black hole and to a white hole.
The manifold description based on this embedding is closely related to the use of Kruskal-Szekeres coordinates, see a remark at the end of \cite{frons}).
This embedding is not asymptotically flat, and it cannot be described in terms of elementary functions.
Note that if we pass to the signature $(+-+---)$, the form (\ref{15}) can also be used
in order to  construct the embedding of a finite region of a manifold with the Schwarzschild metric,
but this embedding will contain singularities.

\subsection{Variant 3: Davidson-Paz embedding}\label{p4var3}
Consider now the cases for which among the terms in $W$ there can be $s=1$.
Using the formula (\ref{5.1}),(\ref{5.2}) we write down all the variants giving the dimension three:
 \disn{18}{
W_0=\{0,1,0,0\}\oplus\{0,0,0,0\}\oplus\{0,0,0,0\},\ns
W_3=\{0,0,0,1\}\oplus\{0,1,0,0\},\qquad
W_4=\{0,1,0,0\}\oplus\{0,0,1,0\}.
\nom}
According to (\ref{z25}),(\ref{z26}) for the variant $W_0$ the initial vector $y_0$ is $D^2_F(r)\oplus B^0(r)\oplus B^0(r)$,
that, according to (\ref{z24}),(\ref{z17.1}),(\ref{z24a1}), gives the embedding function
 \disn{19}{
y^0=h(r)+\hat\g t,\quad
y^1=y^1(r),\quad
y^2=y^2(r)
\nom}
(the three other components follow the eq. (\ref{11}))
where $\hat\g=v\g$ is a dimensionless factor.
We cannot obtain the embedding of the Schwarzschild metric under the form of a surface corresponding to $W_0$,
as it follows from (\ref{19}) that $g_{00}=\pm \hat\g^2=const$.

Consider the variant $W_3$. According to (\ref{z25}),(\ref{z26}) in this case the initial vector $y_0$ is  $B_C^0(r)\oplus D^2_F(r)$,
resulting in the embedding function which, according to (\ref{z24}), (\ref{z17.1}), (\ref{z24a1}), (\ref{z21a2}) for some values of $r$ can be written as
 \disn{20}{
y^0=f(r)\sh(\be t+w(r)),\quad
y^1=f(r)\ch(\be t+w(r)),\quad
y^2=\hat\g t+h(r)
\nom}
(the three other components follow the eq. (\ref{11}))
and for other values of $r$ it can be written under a similar form, but with the interchange of hyperbolic sine and cosine
(depending on the sign of the square of the vector $B^0_C(r)$).
Substituting this embedding function in (\ref{vv1}),(\ref{2}), we found that
with the signature $(+-----)$ the components of the embedding function with $r>r_c\equiv R/(1+\hat\g^2)$
read
 \disn{21}{
y^0=\frac{1}{\be}\sqrt{1+\hat\g^2-\frac{R}{r}}\,\sh(\be t+w(r)),\quad
y^1=\frac{1}{\be}\sqrt{1+\hat\g^2-\frac{R}{r}}\,\ch(\be t+w(r)),
\nom}
while for $r<r_c$ they read
 \disn{22}{
y^0=\frac{1}{\be}\sqrt{\frac{R}{r}-1-\hat\g^2}\,\ch(\be t+w(r)),\quad
y^1=\frac{1}{\be}\sqrt{\frac{R}{r}-1-\hat\g^2}\,\sh(\be t+w(r)),
\nom}
and in both cases $y^2=\hat\g t+h(r)$.
Here the functions $h(r)$ and $w(r)$ can be easily written under the form of certain integrals.
These functions have logarithmic singularities at $r=R$ and $r=r_c$.

This embedding was suggested in \cite{davidson} where one can found an exact expressions for $h(r)$ and $w(r)$.
In order to ensure that the singularity at $r=R$ is due to the choice of coordinates only,
it is sufficient to pass to a new coordinate $t'=t+h(r)/\hat\g$ (similar to the Eddington-Finkelstein one),
because the difference $w(r)-h(r)\be/\hat\g$ has no singularity in this point, see~\cite{davidson}.
In order to remark that the singularity at $r=r_c$ is also of the coordinate type, it is convenient to rewrite the embedding function (\ref{21}) as
 \disn{23}{
\begin{array}{cl}
&\displaystyle y^0=\frac{R}{2\be\sqrt{r_c r}}\ls e^{\be t'+u(r)}-\frac{r-r_c}{R}\,e^{-\be t'-u(r)}\rs,\\
\phantom{\Biggl(}
&\displaystyle y^1=\frac{R}{2\be\sqrt{r_c r}}\ls e^{\be t'+u(r)}+\frac{r-r_c}{R}\,e^{-\be t'-u(r)}\rs,\\
&\displaystyle y^2=\hat\g t',
\end{array}
\nom}
where the function
 \disn{24}{
u(r)=\int\! dr\,\, \frac{1-\sqrt{\frac{4\be^2r_c^2(r-r_c)}{R-r_c}+\frac{r_c^3(R-r)}{r^3(R-r_c)}}}{2(r-r_c)}
\nom}
is smooth at all $r>0$ if $\be^2>2r_c/(27R^2(3R-2r_c))$
(if this condition is violated the radicand will be negative at several $r$).

The manifold defined by the given embedding is everywhere smooth (at $r\to0$ we get  $y^0\to\infty$).
It covers a half of the Riemannian space corresponding to the Schwarzschild solution:
one instance of the region $r>R$ and one instance of the region $r<R$,
related to a black hole (or to a white hole if we change the sign of $y^0$).
In the Kruskal-Szekeres coordinates $v,u$ the covered region is defined by $v+u>0$,
while at $v+u\to0$ we obtain $y^2\to-\infty$.
In the limit $\hat\g\to0$ a part of this manifold passes to the half of the manifold given by the Fronsdal's embedding (\ref{16}).
This embedding is not asymptotically flat, and it cannot be described in terms of elementary functions.
Note that if we pass to the signature $(+-+---)$, the form (\ref{20}) can also be used
in order to  construct the embedding of some part of a manifold with the Schwarzschild metric.

\subsection{Variant 4: asymptotically flat embedding}\label{p4var4}

Consider now the variant $W_4$.
According to (\ref{z25}),(\ref{z26}) in this case the initial vector $y_0$ is $D^2_F(r)\oplus B_A^0(r)$,
that, according to (\ref{z24}),(\ref{z17.1}),(\ref{z24a1}), gives the embedding function
 \disn{25}{
y^0=\hat\g t+h(r),\quad
y^1=f(r)\sin(\al t+w(r)),\quad
y^2=f(r)\cos(\al t+w(r))
\nom}
(the three other components follow the eq. (\ref{11})).
Substituting it in (\ref{vv1}),(\ref{2}), we found that
for the signature $(+-----)$ the embedding function becomes
 \disn{26}{
y^0=\hat\g t+h(r),\ns
y^1=\frac{1}{\al}\sqrt{\frac{R}{r}-1+\hat\g^2}\,\sin(\al t+w(r)),\quad
y^2=\frac{1}{\al}\sqrt{\frac{R}{r}-1+\hat\g^2}\,\cos(\al t+w(r)),
\nom}
where the functions $h(r)$ and $w(r)$ can be easily written under the form of certain integrals.
At $\hat\g=0$ this embedding is a kind of the Kasner's embedding covering the region $r<R$.
On the other hand, if $\hat\g\ge 1$ and $\al$ greater than some value,
one can show that the embedding (\ref{26}) defines a manifold smooth at all $r>0$.

Consider in details a variant of such embedding corresponding to $\hat\g=1$. It  seems to be the most interesting, since the corresponding manifold appears to be asymptotically flat.
In this case
 \disn{27}{
y^0=t+h(r),\quad
y^1=\frac{1}{\al}\sqrt{\frac{R}{r}}\,\sin(\al t+w(r)),\quad
y^2=\frac{1}{\al}\sqrt{\frac{R}{r}}\,\cos(\al t+w(r)),
\nom}
where
 \disn{28}{
h(r)=\int dr \frac{R}{2\al(r-R)}\sqrt{4\al^2+\frac{(R-r)}{r^3}},\quad
w(r)=\int dr \frac{r}{2(r-R)}\sqrt{4\al^2+\frac{(R-r)}{r^3}}.
\nom}
The radicand in the function $h(r),w(r)$ is positive for all $r>0$ if $\al\ge 1/(\sqrt{27}R)$. At $r=R$ these functions have a logarithmic singularity, but it is associated only with the choice of the coordinates.  In order to verify it we can pass to a new coordinate $t'=t+h(r)$ and obtain
 \disn{29}{
y^0=t',\quad
y^1=\frac{1}{\al}\sqrt{\frac{R}{r}}\,\sin(\al t'+u(r)),\quad
y^2=\frac{1}{\al}\sqrt{\frac{R}{r}}\,\cos(\al t'+u(r)),
\nom}
where the function
 \disn{30}{
u(r)=w(r)-\al h(r)=
\frac{1}{2}\int dr \sqrt{4\al^2+\frac{(R-r)}{r^3}}
\nom}
has no singularity at $r>0$.
The integral (\ref{30}) is expressed in terms of elementary functions only at $\al=1/(\sqrt{27}R)$.
In this case the embedding function takes a simple form
 \disn{31}{
\begin{array}{cl}
&\displaystyle y^0=t',\\
{\large\phantom{\Biggl(}}
&\displaystyle y^1=\sqrt{\frac{27R^3}{r}}\,\sin\ls\frac{t'}{\sqrt{27}R}-\sqrt{\frac{(r+3R)^3}{27R^2r}}\rs,\\
&\displaystyle y^2=\sqrt{\frac{27R^3}{r}}\,\cos\ls\frac{t'}{\sqrt{27}R}-\sqrt{\frac{(r+3R)^3}{27R^2r}}\rs\\
\end{array}
\nom}
(the three other components follow the eq. (\ref{11}))
We call "asymptotically flat" the embedding which are constructed in this way.

The manifold defined by the embedding (\ref{29}) at $\al\ge 1/(\sqrt{27}R)$ (and, in particular, the embedding (\ref{31})), is everywhere  smooth.
At $r\to0$ the radiuses of the  helixes defined at fixed $r$ by (\ref{29}) and (\ref{31}), grow infinitely,
i.~e. the central singularity appears to be at the infinity.
Similarly to the Davidson-Paz embedding, the given embedding covers a half of the Riemannian space corresponding to the Schwarzschild solution,
defined by $v+u>0$ in the Kruskal-Szekeres coordinates,
while at $v+u\to0$ it appears that $y^0\to-\infty$.

In contrast to all other symmetric embeddings into the six-dimensional space, this embedding is asymptotically flat.  At $\al=1/(\sqrt{27}R)$, i.~e. under the form (\ref{31}), it is expressed in terms of elementary functions.
An embedding of a very similar structure, also asymptotically flat and expressed in terms of elementary functions, was suggested in work \cite{blashke2010}. But in \cite{blashke2010} the ambient space is seven-dimensional with two time-like directions (i.~e. $R^{2,5}$),
it corresponds to a use of the representation $W=\{0,1,0,0\}\oplus\{0,0,1,0\}\oplus\{0,0,0,0\}$.

Note that if we pass to the signatures $(+++---)$ and $(-++---)$, the  embedding (\ref{25})
can also be used to  construct an embedding of some part of a manifold with the Schwarzschild metric.
Moreover we can construct a global embedding for Reissner-Nordstrom solution if we use the embedding (\ref{25}) with signature $(-++---)$.

\subsection{Variant 5: Fujitani-Ikeda-Matsumoto embedding}\label{p4var5}
Consider now the case when among the terms in the $W$ there can be $s=2$.
Using (\ref{5.2}) we easily get the only variant giving the dimension three:
 \disn{32}{
W_5=\{0,2,0,0\}.
\nom}
According to (\ref{z26}) in this case the initial vector $y_0$ is $B_F^0(r)$,
that, according to (\ref{z24a1}),(\ref{z17.1}), gives us the embedding function
 \disn{33}{
y^+=u(r)+\g t w(r)+\frac{\g^2t^2}{2} f(r),\quad
y^0=w(r)+\g t f(r),\quad
y^-=f(r)
\nom}
(the three other components follow the eq. (\ref{11}))
After transition from light-like coordinates to Lorenz ones
(see the remark after the formula (\ref{z21a3})),
the embedding function reads
 \disn{34}{
y^0=w(r)+\g t f(r),\quad
y^1=\frac{1}{\sqrt{2}}\ls u(r)+\g t w(r)+\ls\frac{\g^2t^2}{2}-1\rs f(r)\rs,\ns
y^2=\frac{1}{\sqrt{2}}\ls u(r)+\g t w(r)+\ls\frac{\g^2t^2}{2}+1\rs f(r)\rs
\nom}
and there are two possibilities of the signature choice: $(++----)$ and $(--+---)$.
Substituting this embedding function in (\ref{vv1}),(\ref{2}),
we found that at $r>R$ the signature must have the first form,
and that the arbitrariness of the constant added to $t$
allows to write the embedding function as
 \disn{35}{
\begin{array}{cl}
&\displaystyle y^0=t\sqrt{1-\frac{R}{r}},\\
{\phantom{\Biggl(}}
&\displaystyle y^1=\frac{1}{\sqrt{2}\,\g}\ls\frac{\g^2t^2}{2}-1\rs \sqrt{1-\frac{R}{r}}+\frac{u(r)}{\sqrt{2}},\\
&\displaystyle y^2=\frac{1}{\sqrt{2}\,\g}\ls\frac{\g^2t^2}{2}+1\rs \sqrt{1-\frac{R}{r}}+\frac{u(r)}{\sqrt{2}},\\
\end{array}
\nom}
where
 \disn{36}{
u(r)=\frac{\g r(2r+3R)}{4}\sqrt{1-\frac{R}{r}}
+\frac{3\g R^2}{8}\ln\ls\frac{2r}{R}\ls 1+\sqrt{1-\frac{R}{r}}\rs-1\rs.
\nom}

This embedding was first suggested in \cite{fudjitani} (where $\g=\sqrt{2}$).
It has also been studied in details in \cite{plebansky95} (where $\g=\sqrt{2}/R$).
The manifold defined in this way
covers only the region $r>R$ and has a conic singularity at $r=R$.
This embedding is not asymptotically flat and is described in terms of elementary functions
(this fact is especially stressed out in \cite{plebansky95}).
Note that if we pass to the signature $(--+---)$, the form (\ref{34}) can also be used
to  construct an embedding of the Schwarzschild metric at $r<R$.

\subsection{Variant 6: an embedding cubic with respect to time}\label{p4var6}
Consider now the case when among the terms in the $W$ there can be $s=3$.
Using the formula (\ref{5.1}) we can easily write down the only variant giving the dimension three:
 \disn{37}{
W_6=\{0,3,0,0\}.
\nom}
It follows from (\ref{5.1}) and (\ref{5.2}) that at $s>3$  there is no variant giving the dimension three,
hence $W_6$ is the last possibility corresponding to a symmetric embedding of the Schwarzschild metric in a flat six-dimensional space.

According to (\ref{z25}) for $W_6$  the initial vector reads $y_0$ is $D^4_F(r)$,
that, according to (\ref{z24}),(\ref{z17.1}), gives the embedding function
 \disn{38}{
y^+\!=h_1(r)+\g t\, h_2(r)+\frac{\g^2t^2}{2} h_3(r)+\frac{v\g^3t^3}{6},\quad
y^2\!=h_2(r)+\g t h_3(r)+\frac{v\g^2t^2}{2},\quad
y^-\!=h_3(r)+v\g t
\nom}
(the three other components follow the eq. (\ref{11})). We see that at $v=0$ this embedding function passes into (\ref{33}),
i.~e. by this way we obtain a generalization of the Fujitani-Ikeda-Matsumoto embedding.
After transition from light-like coordinates to Lorenz ones we obtain
 \disn{39}{
\begin{array}{cl}
&\displaystyle y^0=\frac{1}{\sqrt{2}}\ls h_1(r)+\g t h_2(r)+\ls\frac{\g^2t^2}{2}+1\rs h_3(r)+v\ls\frac{\g^3t^3}{6}+\g t\rs\rs,\\
{\phantom{\Biggl(}}
&\displaystyle y^1=\frac{1}{\sqrt{2}}\ls h_1(r)+\g t h_2(r)+\ls\frac{\g^2t^2}{2}-1\rs h_3(r)+v\ls\frac{\g^3t^3}{6}-\g t\rs\rs,\\
&\displaystyle y^2=h_2(r)+\g t h_3(r)+v\frac{\g^2t^2}{2},\\
\end{array}
\nom}
and there exist again two possibilities of the signature choice: $(-++---)$ and $(+-----)$
(for convenience we changed here the components order and designations compared to (\ref{34})).
Substituting this embedding function in (\ref{vv1}),(\ref{2}),
we found that for the signature $(+-----)$
 \disn{40}{
h_3(r)=\int dr \frac{\sqrt{Rr\ls \frac{R(R-r)}{r^4}+4v^2\g^4\rs}}{2\g(r-R)},\ns
h_2(r)=\frac{1}{2v\g^2}\ls 1-\frac{R}{r}\rs+\frac{h_3(r)^2}{2v},\qquad
h_1(r)=\frac{1}{v}\int dr \ls h'_2(r)h_3(r)-h_2(r)h'_3(r)\rs,
\nom}
where the prime  means the derivative.

The radicand in the function $h_3(r)$ is positive for all $r>0$ if $v\ge \sqrt{27}/(32\g^2 R)$. At $r=R$ the function $h_3(r)$, and hence the function $h_{1,2}(r)$, have a logarithmic singularity.
However, similarly to the cases considered above,
it is due to  the choice of coordinates only.
In order to verify it we can pass to a coordinate $t'=t+h_3(r)/(\g v)$,
and as a result the embedding function reads
 \disn{41}{
\begin{array}{cl}
&\displaystyle y^0=\frac{v(\g t')^3}{6\sqrt{2}}+\frac{\g t'}{\sqrt{2}}\ls\frac{1}{2v\g^2}\ls 1-\frac{R}{r}\rs+v\rs+u(r),\\
{\phantom{\Biggl(}}
&\displaystyle y^1=\frac{v(\g t')^3}{6\sqrt{2}}+\frac{\g t'}{\sqrt{2}}\ls\frac{1}{2v\g^2}\ls 1-\frac{R}{r}\rs-v\rs+u(r),\\
&\displaystyle y^2=\frac{1}{2v\g^2}\ls 1-\frac{R}{r}\rs+v\frac{\g^2t'^2}{2},\\
\end{array}
\nom}
where the function
 \disn{42}{
u(r)=-\frac{1}{2\sqrt{2}v^2\g^3}\int dr \sqrt{\frac{R}{r}\ls \frac{R(R-r)}{r^4}+4v^2\g^4\rs}
\nom}
is smooth at all $r>0$ if $v\ge \sqrt{27}/(32\g^2 R)$.
We call this embedding "cubic with respect to time".

The manifold defined by this embedding is everywhere smooth
(at $r\to0$ either $y^0\to\infty$, or $y^1\to\infty$).
Similarly to the embeddings described in Sections \ref{p4var3}, \ref{p4var4},
this embedding covers a half of the Riemannian space corresponding to the Schwarzschild solution,
this space being defined by the condition $v+u>0$ in the Kruskal-Szekeres coordinates,
while at $v+u\to0$ it appears that $y^0\to-\infty$.
The given embedding is not asymptotically flat
and it cannot be described in terms of elementary functions.
Note that if we pass to the signature $(-++---)$, the form (\ref{39}) can also be used
in order to  construct an embedding of some part of a manifold having the Schwarzschild metric.

\section{Conclusion}
Using the method to construct surfaces with a given symmetry suggested in Section \ref{p2}
we succeeded to show that there exist six kinds of embeddings of the Schwarzschild metric in a flat six-dimensional space if the embeddings  have  the symmetry of Schwarzschild solution.
Four of them are known from the works \cite{kasner3,frons,davidson,fudjitani}, while the two others are new.
The only asymptotically flat embedding in a six-dimensional space appears to be the new embedding considered in Section \ref{p4var4}.

Various kinds of embeddings differ,  first of all, by the method of  realization  of the symmetry corresponding to the shift of the $t$ coordinate.
For Kasner's and Fronsdal's embeddings this shift is equivalent to a rotation and a pseudo-rotation (i.~e. Lorenz boost) in the ambient space, respectively.
For the asymptotically flat embedding from Section \ref{p4var4} and for Davidson-Paz embedding
it is equivalent to a rotation and a pseudo-rotation,
associated with the simultaneous shift in the direction orthogonal to the plane of rotation.
For Fujitani-Ikeda-Matsumoto embedding
the shift $t$ is equivalent to a rotation in a light-like plane,
and for the cubic with respect to time embedding from Section \ref{p4var6} --
to a similar rotation associated to a special kind of shift.

The method suggested in the present work can be used to  construct the embeddings of the Riemannian spaces  which  have enough high symmetry.

\vskip 0.5em
{\bf Acknowledgments.}
The authors are grateful to the Prof. M.V.~Ioffe and Prof. V.D.~Lya\-khovsky for useful discussions.
The work of one of the authors (A.~A.~Sh.) was supported by the non-profit Dynasty Foundation.

\end{document}